\documentclass[aps,twocolumn]{revtex4}

\usepackage{graphicx}
\usepackage{dcolumn}
\usepackage{bm}

\begin{document}

\title{Ground-state characterization of Nb charge-phase Josephson
qubits}

\author{H. Zangerle, J. K\"onemann, B. Mackrodt, R. Dolata, S. V. Lotkhov, S. A. Bogoslovsky, M. G\"otz, and A.B. Zorin}
 \affiliation{Physikalisch-Technische Bundesanstalt, Bundesallee 100, 38116 Braunschweig, Germany}%

\date{\today}

\begin{abstract}
We present investigations of Josephson charge-phase qubits of
SQUID-configuration inductively coupled to a
radio-frequency-driven tank circuit enabling the readout of the
states by measuring the Josephson inductance of the qubit. The
circuits including junctions with linear dimensions of $60$ nm
$\times$ $60$ nm and $80$ nm $\times$ $80$ nm are fabricated from
Nb/AlO$_{\rm x}$/Nb trilayer and allow the determination of
relevant sample parameters at liquid helium temperature. The
observed partial suppression of the circulating supercurrent in
the qubit loop at 4.2 K is explained within the framework of a
quantum-statistical model. We have probed the ground-state
properties of qubit structures with different ratios of the
Josephson coupling to Coulomb charging energy at 20 mK,
demonstrating both the magnetic control of phase and the
electrostatic control of charge of the qubit.
\end{abstract}

\maketitle

\section{\label{sec:level1} Introduction}
\begin{figure}
\includegraphics[width=0.9\linewidth]{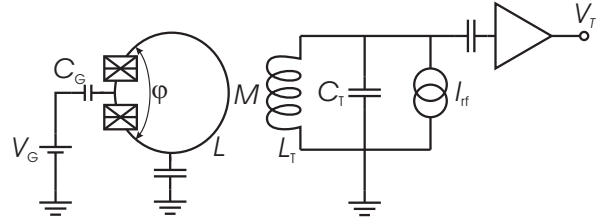}
\caption{\label{fig:fig2} Scheme of the experiment. The core
element  is a double Josephson junction (two crossed boxes) with a
capacitive gate coupled to its island, i.e. the Bloch transistor,
embedded in a macroscopic superconducting loop. The loop is
inductively coupled to an rf-driven tank circuit connected to a
cold preamplifier. The capacitance between the loop and the ground
is assumed to be much larger than the total capacitance of the
island.}
\end{figure}

Superconducting structures with mesoscopic Josephson tunnel
junctions can provide a basis for electronic devices operating on
single Cooper pairs. Prominent examples are the superconducting
quantum bit (qubit) circuits which are regarded as  promising
elements for a scalable quantum computer \cite{makhklin01}. The
Josephson charge-phase qubit \cite{cpqb1} is based on a
Cooper-pair box \cite{qpbox} of SQUID-configuration, i.e. a
superconducting loop interrupted by two small-capacitance
junctions with an island  in between, which is capacitively
coupled to a gate electrode (i.~e.~the Bloch transistor
\cite{q-SG}). The charging energy $E_{\rm C}$ and the Josephson
coupling energy $E_{\rm J}$ are typically of the same order, so
the dimensionless parameter $\lambda = E_{\rm J}/E_{\rm C}$ is of
the order of one. Moreover, our circuit comprises both the qubit
and its readout \cite{cpqb2}. The transistor can be operated as a
box (qubit) whose distinct quantum states with energies $E_n, n
=0,1,2,...,$ are associated with different Bloch-bands of the
system \cite{sqc}. The eigenfunctions $|n,q\rangle$ are the Bloch
wave functions of a particle in the periodic (Josephson)
potential. Here $n$ is the band number and $q$ the quasicharge
governed by the gate voltage $V_{\rm G}$, i.~e.~$q=C_{\rm G}V_{\rm
G}$, where $C_{\rm G}$ is the gate capacitance, see, e.g.,
Ref.\,\cite{zorin96}. The quantum states of the transistor also
 involve the phase coordinate $\varphi$ set by the external
magnetic flux $\Phi_{\rm dc}$ applied to the SQUID loop. The
variable $\varphi$ behaves almost classically and is regarded as a
parameter. Due to two control parameters (charge and phase), the
eigenenergies $E_{\rm n}(q,\varphi)$, the transition frequency
$\nu_{10} =(E_1-E_0)/h$ as well as - in the case of multiple
qubits - the strength of mutual coupling, can be varied in the
wide range. The read-out of our qubit can be performed similar as
in the rf-SQUID-based impedance measurement technique pioneered by
Rifkin and Deaver \cite{rifkin76}: the qubit eigenstates can be
distinguished by the Josephson inductance of the effective weak
link, i.e. the transistor, included in the loop whose impedance is
probed by small rf-oscillations induced by an inductively coupled
resonant tank circuit \cite{cpqb2,zorin03}, see the equivalent
circuit in Fig. \ref{fig:fig2}.

Recently, Born {\it et al.} \cite{imt2} have confirmed the
aforementioned read-out conception in the spectroscopic experiment
with Al qubit. The authors demonstrated the detection of
microwave-power-induced interband transitions by tracking the
resonant response of the tank. However, due to the rather large
ratio $\lambda=E_{\rm J}/E_{\rm C} \approx 30$, manipulation of
the qubit state via its gate was in their experiment limited to a
narrow (about 5\%) interval of values of the flux. In this paper,
we present a comprehensive mapping of the ground state of Nb
qubits operating in the charge-phase regime over a wide range of
the parameters $q$ and $\varphi$. This is done for the most
illustrative case of the values of the parameter $\lambda \sim 1$.
In this regime, the behaviour of the system strongly depends on
the value of $\lambda$. Moreover, the effect of thermal
fluctuations is studied comparatively at 20\,mK and 4.2\,K for the
same samples. These investigations strongly benefit from the
advanced technology for the fabrication of sub--$\mu$m Josephson
junctions based on Nb/AlO$_{\rm x}$/Nb trilayers with a large
superconducting energy gap $\Delta$ and a low sub-gap leakage
current \cite{dolata05}. Applying this technology allows all
inductive components of the circuitry shown in Fig. \ref{fig:fig2}
to be integrated on one chip.

The material presented in this paper is organized as follows. In
Sec. II, we briefly outline the considerations that motivated our
choice of experimental parameters. In Sec. III, radio-frequency
measurement results for niobium-based Bloch transistors will be
presented and quantitatively analyzed. Especially the
fluctuation-induced partial suppression of the critical current
through Bloch transistors will be discussed within a
quantum-statistical model. In Sec. IV, the concluding remarks will
be made.

\section{\label{sec:level3} Design of experiment}\label{exp}

\subsection{Parameters of the samples}


The operation of a charge-phase qubit bases on the single charge
phenomena thus requiring the charging energy $E_{\rm C}= e^2/2C$
of the small island electrode ($C=C_{\rm J1}+C_{\rm J2}+C_{\rm G}$
denotes the corresponding total capacitance including the
capacitances of the junctions $C_{\rm J1,J2}$) to be large,
compared to the thermal energy $k_{\rm B}T$. In the case of our Nb
technology the parameter $E_{\rm C}\approx 50-80\,\mu$eV roughly
corresponds to temperature $\sim 1$\,K. The rather balanced
situation with the target value $\lambda\sim 1$, i.e. the
Josephson coupling energy of individual junctions $E_{\rm J0} =
(\Phi_0/2\pi)I_{\rm c0} \approx 100\,\mu$eV, where $\Phi_0=h/2e
\approx 2.07\cdot10^{-15}$\,Wb is the magnetic flux quantum,
corresponds to nominal critical currents of the individual
junctions $I_{\rm c0}$ of several tens of nA \cite{comment1}.




The Bloch transistor is included in a double superconducting loop
of octagon shape with  an outer dimension of 1.0\,mm (see
Fig.\,2). Such a gradiometer design improves the stability against
homogeneous magnetic field noise. The total inductance value of
such a loop $L \approx 0.7$\,nH was estimated using Mohan's
formula \cite{mohan}. This value is sufficiently small in the
sense that, firstly, the corresponding magnetic energy is rather
large, i.e. $E_M = \Phi_0^2/2L \approx 20$\,meV and, secondly, the
dimensionless screening parameter, i.e. $\beta_L \equiv 2\pi
LI_{c0}/\Phi_0 \ll 1$, is small. The former relation ensures the
suppression of flux fluctuations and fixes the phase $\varphi$
across the transistor \cite{klus}. The latter one ensures the
required non-hysteretic operation regime in the case of a
single-junction SQUID circuit \cite{hansma,rifkin76}, as well as
of the realistic qubit circuit having finite asymmetry in the
critical currents of individual junctions \cite{zorin03}.
Moreover, the total inductance of the closed loop circuit is
determined mostly by the transistor Josephson inductance. This
inductance is directly related to the local curvature of the
surface $E_n(q,\varphi)$ taken for fixed $q$ \cite{zorin03}, i.e.
\begin{equation}
L_{\rm J}(n,q,\varphi)=
\left(\frac{\Phi_0}{2\pi}\right)^2\cdot\left[\frac{\partial^2E_{\rm
n}(q,\varphi)}{\partial\varphi^2}\right]^{-1}\label{curve}.
\end{equation}
In our measurements of the circuit in the ground state the band
index $n=0$, so we introduce the notation $L_{\rm J}(0,q,\varphi)
\equiv L_{\rm J}(q,\varphi)$.


The loop is coupled through the mutual inductance $M$ to the coil
of a resonant tank circuit formed by the double octagon-shaped
spiral inductor with total inductance $L_{\rm T}=0.15-0.2\,\mu$H
(see Fig.\,2a) and the capacitance of the coaxial cable connecting
the tank to the preamplifier. The bare resonance frequency of such
a tank circuit with quality factor $Q\approx 250$ is $f_0 \approx
70$\,MHz. Due to the coupling to the qubit, the effective
inductance of the circuit is changed \cite{hansma},
\begin{equation}
L_{\rm eff} = L_{\rm T}-M^2L_{\rm J}^{-1}(q,\varphi),
\end{equation}
in a gate-charge and phase specific way. The resulting shift
$\Delta f$ of the resonant frequency, is \cite{zorin03}
\begin{equation}
\frac{\Delta f}{f_0}=-\frac{1}{2}k^2\beta_{\rm L}\frac{\Phi_0/2\pi
I_{\rm c}}{L_{\rm J}(q,\varphi)}.\label{detuning}
\end{equation}
Here, $k=M/\sqrt{LL_{\rm T}} < 1$ denotes the coupling coefficient
which is determined by the arrangement of the inductively coupled
conductors. The integrated on-chip design of our qubit allows
close mutual arrangement of the loop and the coil and, therefore,
rather large values of $k$. The tank circuit is driven by a
combined current signal, consisting of a dc part fixing the
working point, and an rf component $I_{\rm rf}$ at the frequency
$f$ close to $f_0$.


\begin{figure}
\includegraphics[width=0.9\linewidth]{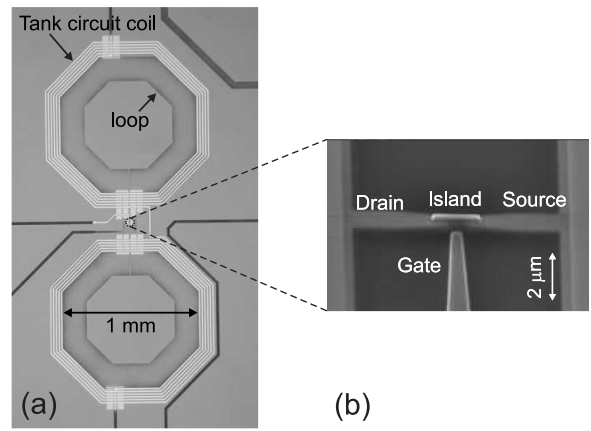}
\caption{\label{fig:fig1s} (a) The image of the sample of the gradiometer
configuration with the Bloch-transistor structure in the center. (b) The
zoomed-in microphotograph of the transistor.}
\end{figure}


The cold preamplifier is based on an AGILENT ATF-10136 GaAs field-effect
transistor. The output of the preamplifier is connected by the coaxial cable to
the room-temperature amplifier. The overall gain of this two-stage scheme is
about 26\,dB in the range of typical resonant frequencies up to 90\,MHz.
The amplified rf voltage across the tank is fed to a lock-in
amplifier measuring both the amplitude $V_{\rm T}$ and the phase
shift $\alpha$ relative to the reference rf signal.

\subsection{Pre-characterization of Bloch transistors}

\begin{figure}
\includegraphics[width=0.99\linewidth]{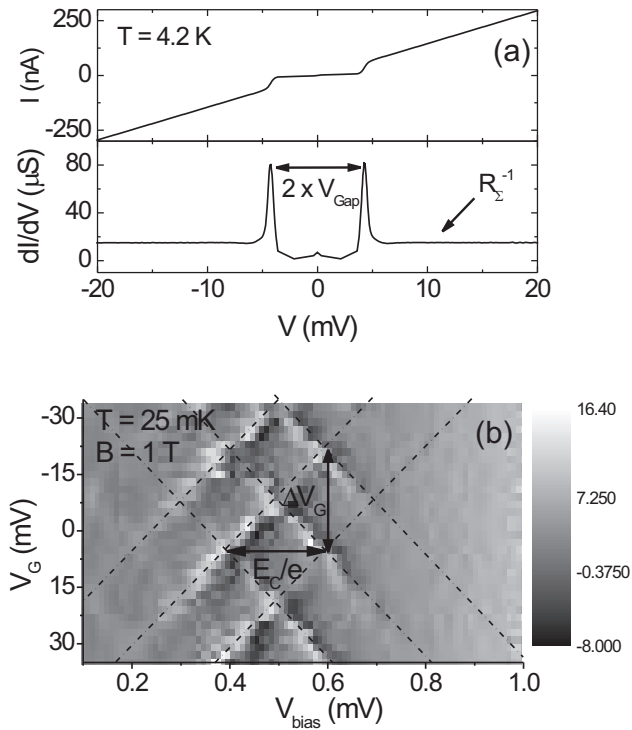}
\caption{\label{fig:fig2s} (a) Typical dc \textit{I-V} curve of
the stand-alone test transistor from wafer B  at 4.2 K. The
superconducting gap-voltage in this sample is $V_{\rm Gap}=4.17$
mV and the normal-state resistance $R_{\Sigma}=67$ k$\Omega$ of
the series double-junction (transistor). (b) Contour-plot of the
differential conductance $dI/dV$ in $\mu$S of an isolated test
transistor of wafer B as a function of gate  voltage and bias
voltage measured at 25 mK.}
\end{figure}

Embedding the Bloch transistors in a superconducting loop,
unfortunately, renders its pre-characterization in a simple dc
configuration impossible. To get nevertheless an estimate of the
relevant parameters, i.e. the Josephson coupling energy $E_{\rm
J0}$ and the charging energy $E_{\rm C}$, we characterized the
similar stand-alone Bloch transistor structures fabricated from
the same wafer, hence made from the same Nb/AlO$_x$/Nb trilayer,
and having the same dimensions. From $I$-$V$ measurements at 4.2 K
(shown in Fig. \ref{fig:fig2s}a) we extract the sum asymptotic
resistance $R_{\Sigma}$ and the gap voltage $V_{\rm Gap}$.
Resistance $R_{\Sigma}$ is assumed to equal twice the normal state
resistance $R_{\rm N} \approx 33.5\,\textrm{k}\Omega$ of one from
two nominally identical junctions while the superconductor energy
gap $\Delta=eV_{\rm Gap}/4 \approx 1.04$\,meV. Inserting these
values into the Ambegaokar-Baratoff relation $R_{\rm N} I_{\rm
c0}\approx (\pi/2)\Delta/e$ \cite{ambega}, we estimate the
critical current $I_{\rm c0}$ and, finally,  $E_{\rm J0}$, the
Josephson coupling energy of one junction as listed in Table I.
Note that $I_{\rm c0}$ values of 25 nA (wafer A) respectively 45
nA (wafer B) differ almost by a factor of 2. The radio-frequency
measurements performed at 4.2\,K with the single junctions
inserted instead of the transistors in the similar loops (see
Fig.~\ref{fig:fig2}), i.e. the rf-SQUID configuration, gave almost
similar values for $I_{\rm c0}$.

The values of $E_{\rm C}$ were derived from the gate and bias
voltage dependencies of the current of the stand-alone transistors
at 20\,mK and perpendicular magnetic field of up to 2\,T. This
magnetic field partially suppresses the superconductivity of the
Nb electrodes and enhances the single-electron tunneling at small
voltage bias \cite{pavolotsky}. For the bias below the gap voltage
we found the characteristic diamond-like pattern resulting from
Josephson quasiparticle cycles \cite{tinkham}; the bias voltage
period provides an estimate of the charging energy (see Fig.
\ref{fig:fig2s}b and Table \ref{tab:tabn}), while the gate voltage
period $\Delta V_{\rm G}\approx 28$ mV gives the value of the gate
capacitance, $C_G=e/\Delta V_G\approx 6$\,aF.


\begin{table}[!h]
\caption{\label{tab:tabn} Parameter overview for the experiments with the Bloch
transistor samples T1 and T2.
}
\begin{ruledtabular}
\begin{tabular}{lcc}
\mbox{ } & T1 ($60$ nm $\times$ $60$ nm) & T2 ($80$ nm $\times$ $80$ nm) \\
\hline  $E_{\rm J0}$ ($\mu$eV)  & 50 & 95  \\
  $E_{\rm C}$ ($\mu$eV)  & 80 & 45 \\
  $L$ (nH) & 0.7 & 0.7  \\
  $L_{\rm T}$ ($\mu$H) & 0.20 & 0.15\\
 $M$ (nH)& 4.6 & 3.8  \\
 $k$ & 0.42 & 0.39  \\
 $I_{\rm c}$ (nA) & 3.9 & 17  \\
 $\beta_{\rm L}$ & 0.008 & 0.036 \\
 $f_0$ (MHz)& 68.80 & 76.18  \\
 $Q$ & 250 & 235  \\
\end{tabular}
\end{ruledtabular}
\end{table}
\section{Radio-frequency measurements}

\subsection{Resonance curves}
\begin{figure}
\includegraphics[width=1\linewidth]{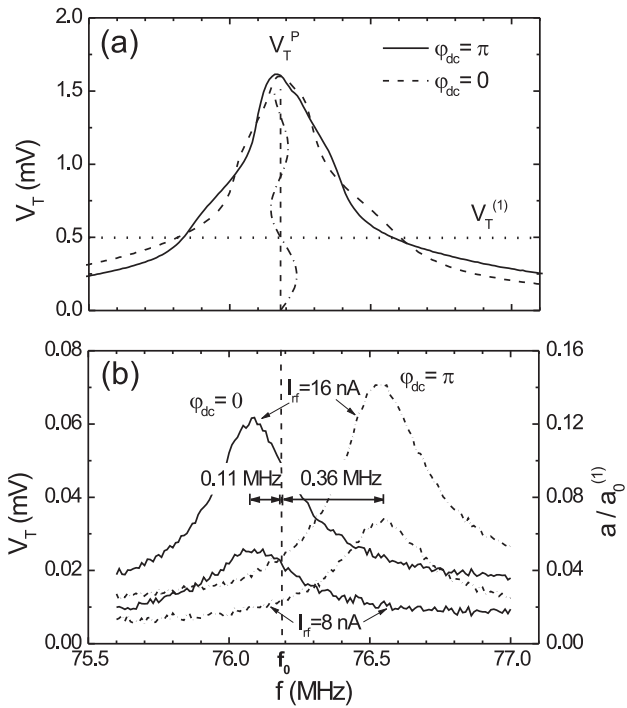}
\caption{\label{fig:fig3}(a) Typical resonance curves for sample
T2 at $20$ mK. Shown is the detected tank-circuit voltage as a
function of the applied frequency of the rf-excitation for a
driving-current of $I_{\rm rf}\approx 320$ nA. The dashed-dotted
line shows the amplitude-dependency $f(a)$ of the
resonance-frequency for an assumed harmonic phase-dependence, as
described by Eq.~(\ref{wa}). (b) Typical resonance curves  of
sample T2 for a low excitation power, i.e. driving-currents
$I_{\rm rf}\approx 8$ nA and $I_{\rm rf}\approx 16$ nA.}
\end{figure}

First, the resonance measurements of the Bloch transistors were
performed in order to calibrate the rf oscillation amplitude of
the phase, $\varphi(t)=\varphi_{\rm dc}+a\sin{(2\pi ft)}$, with
$a$ being proportional to the amplitude of the rf-flux
oscillations in the loop, $\Phi_a=a\Phi_0/2\pi$. The resonance
curves $V_{\rm T}-f$ for a moderate pumping amplitude are shown in
Fig.\,\ref{fig:fig3}a. The curves exhibit characteristic "nodes"
typical of rf-SQUIDs \cite{shnyrkov}, where, at certain pumping
levels, the sensitivity of the tank oscillation amplitude against
the dc bias, i.e. the stationary Josephson phase difference
$\varphi_{dc}$, disappears. Assuming a sinusoidal current-phase
relation (CPR), these "nodes" are related to the
amplitude-dependent resonant frequency for fixed $\varphi_{\rm
dc}$ given by the formula \cite{likha}:
\begin{equation}
 f(a)=f_0\cdot\left\{1+k^2\beta_{\rm L} [J_1(a)/a]
 \cos{\varphi_{\rm dc}}\right\}\label{wa}
\end{equation}
 with $J_1(a)$ denoting the first order
Bessel-function. If $a$ equals one of the positive zeros
$a_0^{(i)}$ ($i=1,2,…$) of function $J_1$, the resonant frequency
is no longer dependent on $\varphi_{dc}$ or on the dc bias
current.
The amplitude $V_{\rm T}^{(1)}$ for the first node can be assigned
to the respective value $a_0^{(1)}$, thus providing calibration
according to $a=3.83 V_{\rm T}^{p}/V_{\rm T}^{(1)}$, where $V_{\rm
T}^{p}=\max V_{\rm T}$. The good agreement between the calculated
$f(a)$-dependence (dash-dotted line in Fig. \ref{fig:fig3}a) and
the position of the nodes illustrates, that the transistor sample
investigated behaves for rather large amplitudes approximately as
an element with harmonic CPR. In other words, the contribution of
higher harmonics in the CPR to the resulting resonance frequency
$f(a)$ is small.

For the small pumping levels, the resonance curve  is shifted as a
whole when changing the dc bias current, see
Fig.\,\ref{fig:fig3}b. In contrast to the situation for the large
amplitude of oscillations, the non-harmonic shape of the CPR
 plays  a crucial role here as expressed by Eq.
(\ref{curve}) relating the local curvature of the band with the
Josephson inductance. The frequency shift determined for the two
opposite values of the external dc magnetic flux applied to the
loop is no longer symmetric relative to the tank-circuit frequency
since the reverse inductance $L_{\rm J}^{-1}(q,\varphi)$ producing
this shift (see Eq.\,(3)) is not exactly proportional to
$\cos\varphi$ here.
This asymmetric splitting appears so clear in our experiment
because of the rather large value of the product $k^2Q\beta_{\rm
L}\approx 1.6$ for sample T2 (cf. experiments on rf-SQUIDs with
large values of this product in Ref.\,\cite{dmitrenko}). The
minimum rf-excitation which allows us to work with a sufficiently
high signal-to-noise ratio in our measurements corresponds to
values $a\approx 0.1 a_{0}^{(1)}$.

For a sufficiently small value of $a$ the phase angle $\alpha$
between the driving signal $I_{\rm rf}$ and the voltage
oscillations is given by the formula
\begin{equation}
\tan{\alpha}=2Q(\Delta f/f_0-\xi), \label{fa}
\end{equation}
with the resonance frequency detuning $\Delta f$ resulting from
the transistor's Josephson inductance Eq.\,(\ref{detuning}) and
the relative shift $\xi=(f-f_0)/f_0$ of the operation frequency
$f$. The measurement of this angle for different values of the dc
flux and gate charge at a sufficiently low temperature allows
 the curvature of the qubit ground state surface to be mapped.

\subsection{Radio-frequency measurements at 4.2 K}

As long as our samples are fabricated from Nb films having a
critical temperature of about 9\,K, they preserve the Josephson
properties and can be measured at a temperature of $T=4.2$\,K.
Earlier, the rf measurements of transistors which comprise
somewhat larger Nb junctions (with dimensions down to 300\,nm by
300\,nm) and are included in low-inductance loops were
successfully carried out at 4.2\,K by Il'ichev et al. \cite{imt1}.
These samples showed a clear dependence of the phase angle on the
applied dc flux and allowed  the critical current of the junctions
to be evaluated (about 55\,nA). Due to the large capacitance of
the island in these samples, the charging energy was small (about
4\,$\mu$eV, i.e. 50\,mK) and the values of parameter $\lambda$
were rather large ($> 30$), so these transistors behaved
classically at an operation temperature of 4.2\,K, i.e. like two
classical Josephson junctions connected in series. Surprisingly,
our qubit samples with significant charging energy, i.e. $\lambda
\sim 1$, also exhibited a clear dependence of $\alpha$ on the dc
flux at this temperature.

\begin{figure}
\includegraphics[width=0.94\linewidth]{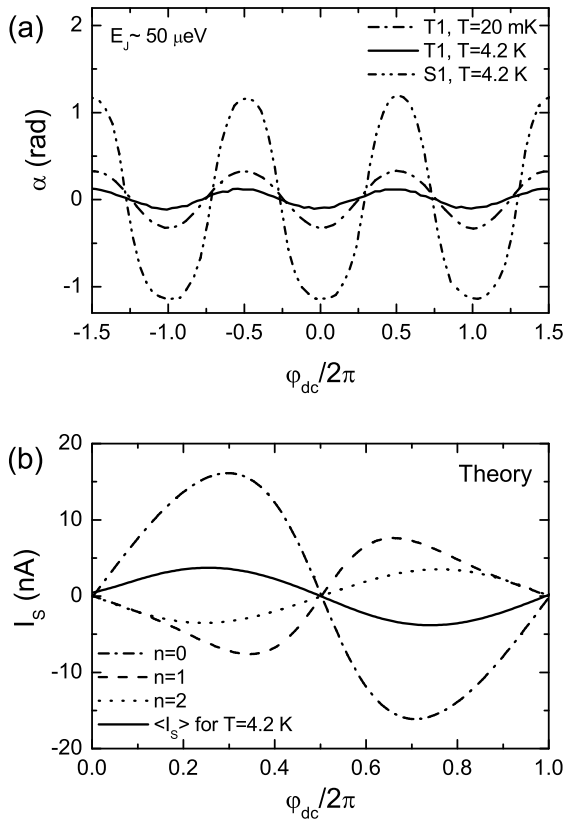}
\caption{\label{fig:fig4} (a) Phase-shift $\alpha$  as a function
of the phase ($\varphi_{dc}\propto \Phi_{dc}$) for a
single-junction (dash-double-dotted line) of wafer A at 4.2 K and
for the double-junction sample $T1$ of wafer A at 20 mK
(dash-dotted line) and 4.2 K (solid line). The nominal value of
the Josephson energy of each junction of wafer A is about 50
$\mu$eV. (b) Current-phase relations for the ground-band
(dashed-dotted line), the first (dashed line) and second band
(dotted line) of the Bloch-transistor calculated for $E_{\rm
J0}=100$ $\mu$eV, $E_{\rm C}=45$ $\mu$eV and $q=0.5$. The thick
solid line indicates the observable value of the supercurrent
$I_\textrm{S}(\varphi,q)$ derived from Eq. (\ref{average2}).}
\end{figure}

Figure\,\ref{fig:fig4}a shows the periodic dependencies
$\alpha$-$\varphi_{\rm dc}$ measured in the qubit sample T1 at
4.2\,K (solid line) and 20\,mK (dash-dotted line). For comparison,
the corresponding curve (measured at 4.2\,K) for the similar
single Josephson junction included in the identical rf circuit is
shown
by a dash-double-dotted line. 
This plot demonstrates first the reduction of the critical current
of the transistor in the ground state at $T=20$\,mK ($k_BT \ll
E_{\rm J},E_{\rm C}$) in comparison to a single junction. This is
due to the charging effect of the island (see, for example, the
experiments \cite{chargeexp,chargeexp2}). Secondly, one can see a
further reduction of the critical current observed at the elevated
temperature of 4.2\,K.

Such behavior can be explained by a simple model which takes the
mixed Bloch states at values of thermal energy $k_BT$ into account
which are comparable to the interband energies. In this case the
observable supercurrent $\langle \hat{I}_{\rm S}\rangle$ is found
as the quantum-statistical averaging over the canonical ensemble,
\begin{equation}\label{average}
I_{\rm S}(\varphi,q) = \langle \hat{I}_{\rm S} \rangle =
\textrm{Tr}(\hat{I}_{\rm S} e^{\hat{H}/k_{\rm
B}T})/\textrm{Tr}(e^{\hat{H}/k_{\rm B}T}),
\end{equation}
where $\hat{H}$ denotes the Hamiltonian of the total system
including the electromagnetic environment in the thermal
equilibrium at a temperature $T$. In view of the small gate
capacitance ($C_{\rm G} \ll e^2/2k_{\rm B}T$) and the small loop
inductance ($L<(\Phi_0/2\pi)^2/k_BT$), the variables $\varphi$ and
$q$ can be regarded as classical parameters. The full expression
for the supercurrent operator $\hat{I}_{\rm S}$ was derived in
Ref.\,\cite{zorin03}. The diagonal matrix elements contributing to
the expectation value entering in Eq.\,(\ref{average}) are,
therefore, equal to
\begin{equation}\label{I_S}
\langle n|\hat{I}_{\rm S}|n \rangle = \frac{2\pi}{\Phi_0}\frac{E_{\rm J1}E_{\rm
J2}}{E_{\rm J}(\varphi)} \sin \varphi \,\langle n|\cos{\chi}|n \rangle,
\end{equation}
where the effective Josephson coupling energy of the transistor is
given by $E_{\rm J}(\varphi)=(E_{\rm J1}^2+E_{\rm J2}^2+2E_{\rm
J1}E_{\rm J2}\cos\varphi)^{1/2}$ and $\chi$ is the operator of the
transistor island's phase (conjugate to the operator of the
island's charge). Note that the value $|\langle n|\cos\chi|n
\rangle |< 1$ (see the plots of this matrix element for $n=0$ in
Ref.\,\cite{zorin96}) also depends on the phase $\varphi$, because
$\varphi$ determines the ratio $\lambda=E_{\rm J}(\varphi)/E_{\rm
C}$.

Note that due to the different signs of the term $\langle
n|\cos{\chi}|n \rangle$ the supercurrent in the different Bloch
bands $\langle n|\hat{I}_{\rm S}|n \rangle$, $n=0, 1, 2,...$ has
usually inverted the phase dependencies (see the calculated curves
in Fig.~\ref{fig:fig4}b as well as Fig.~\ref{fig:fig2s} in
Ref.\,\cite{cpqb2}). These anti-phase terms contribute with the
corresponding Boltzmann's factors to the observable value
Eq.\,(\ref{average}). Finally, the supercurrent value to be
detected in the 4.2~K measurement is equal to
\begin{eqnarray}\label{average2}
I_{\rm S}(\varphi,q)& = &(2\pi/\Phi_0)E_{\rm J1}E_{\rm J2}\sin \varphi
/E_{\rm J}(\varphi) \nonumber\\
&\times & \sum_{n=0}^N \sum_{m=0}^1 \langle n|\cos{\chi}|n
\rangle e^{-E_{\rm n}(q+me,\varphi)}\nonumber\\
& / &\sum_{n=0}^N \sum_{m=0}^1 e^{-E_{\rm n}(q+me,\varphi)}.
\end{eqnarray}
The summation over $m$ takes into account both even and odd
configurations of charge on the island. Both of these
configurations are realized due to the unavoidable single electron
tunneling at the elevated temperature. An analysis of
Eq.\,(\ref{average2}) shows that taking into account the three
lowest energy bands, i.e. choosing $N=2$, is sufficient for an
adequate description of our experiments with the given sample
parameters and temperatures up to $T = 4.2$\,K. Then, higher bands
($n > 2$) are sparsely populated and do not essentially contribute
to $I_{\rm S}$. As a tendency, the dependence of $I_{\rm
S}(\varphi,q)$ on the gate charge $q$ practically vanishes as soon
as higher bands get involved. The phase dependence becomes almost
harmonic.

In Table \ref{tab:ta1} we compare the critical-current values
calculated according to our model, $I_{\rm c}^{\rm (th)}$, with
those extracted from the measurements, $I_{\rm c}^{\rm (m)}$, for
both our transistor samples at 20 mK and 4.2 K. From the
experimental data, we obtained the critical currents according to
the relation
\begin{equation}
I_{\rm c}^{\rm (m)}=
\frac{\Phi_0}{2\pi}\cdot\frac{\tan(\delta\alpha/2)a}{k^2QLJ_1(a)},
\end{equation}
following from Eqs.~(\ref{detuning}) and (\ref{fa}). Here we
introduce the value $\delta\alpha=\max_{\{\Phi_{dc}\}}\alpha -
\min_{\{\Phi_{dc}\}}\alpha$. For both temperatures considered here
we find that the theoretical and the experimental critical
currents are in good agreement. Moreover, the fact that $I_{\rm
c}^{\rm (th)}$ does not differ for zero temperature and 20 mK,
indicates, that measurements at the base temperature of the
dilution refrigerator explore indeed the ground state.

\begin{table}
\caption{\label{tab:ta1} Theoretical and experimental values of
the critical currents of two Bloch transistor samples at a
sufficiently low temperature of 20 mK and at 4.2 K. For the
$I_{\rm c}^{\rm (th)}$ calculation based on the parameters $E_{\rm
J0}$ and $E_{\rm C}$ listed in Table I, $E_{\rm J0}=E_{\rm
J1}=E_{\rm J2}$ is assumed as justified by the weak influence of
the imbalance parameter $b=(E_{\rm J1}-E_{\rm J2})/(E_{\rm
J1}+E_{\rm J2})$ found in the simulations.}
\begin{ruledtabular}
\begin{tabular}{lcc}
\mbox{ } & T1 & T2\\
\hline $I_{\rm c}^{\rm (th)}$ at $T=20$ mK & 4.5 nA & 16 nA\\
$I_{\rm c}^{\rm (m)}$ at $T=20$ mK & 3.9 nA & 17 nA\\
\hline
$I_{\rm c}^{\rm (th)}$ at $T=4.2$ K & 0.9 nA & 4.3 nA\\
$I_{\rm c}^{\rm (m)}$ at $T=4.2$ K & 0.6 nA & 3.0 nA\\
\end{tabular}
\end{ruledtabular}
\end{table}

\subsection{Mapping of the ground state}

\begin{figure}
\includegraphics[width=1.1\linewidth]{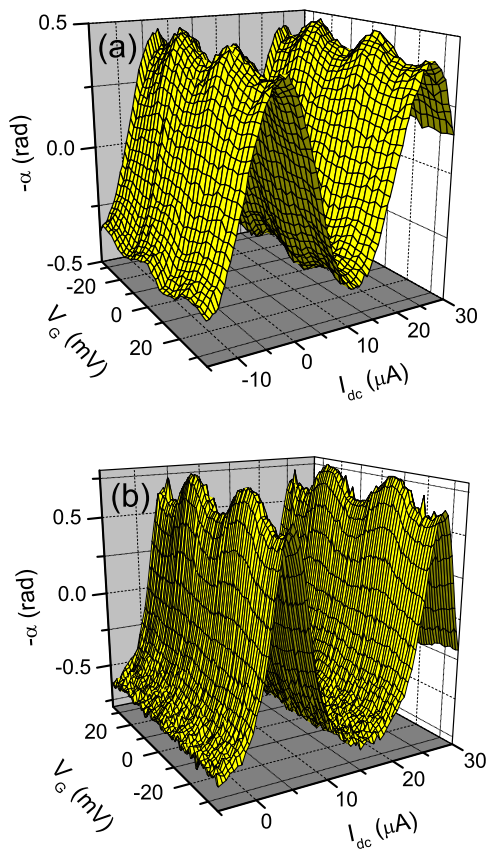}
\caption{\label{fig:fig5} Phase-shift $\alpha$  as a function of
the applied dc-flux ($\propto I_{\rm dc}$) and gate voltage
$V_{\rm G}$ at $T=20$~mK for samples (a) T1 and (b) T2.}
\end{figure}

Figure \ref{fig:fig5} presents the phase dependence as a function of both
external dc-flux (corresponding to the Josephson phase $\varphi$) and the
gate-voltage $V_{\rm G}$ proportional to the quasi-charge $q$ for the samples
T1 and T2.
A nearly sinusoidal dependence of the phase $\alpha$ on
$\varphi_{\rm dc}$ is to be seen which is modulated periodically
by the applied gate voltage $V_{\rm G}$.

A closer look at Fig.\ \ref{fig:fig5} reveals that the amplitude
of the oscillations of $\alpha$ and, therefore, the critical
current is smaller in T1 than in T2, as expected from the
Ambegaokar-Baratoff-values of $E_{\rm J}$ and also because of the
smaller $\lambda$-ratio, due to the charging effects. The nominal
values $E_{\rm J0}$ differ by less than a factor of 2, whereas the
critical currents differ by more than a factor of 4. Related to
that, the dc-bias dependence of $\alpha$ is closer to the
sinusoidal one for sample T1 having the smaller value of
$\lambda=0.7$ \cite{Zorin3}.

\begin{figure}
\includegraphics[width=.95\linewidth]{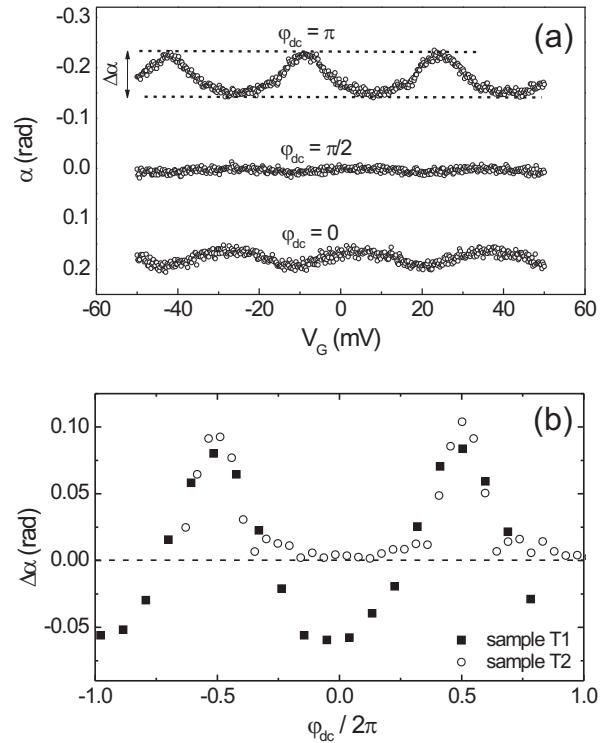}
\caption{\label{fig:fig4s} (a) Gate modulation-curves
$\alpha(V_{\rm G})$ of T1 for three different values of the
applied magnetic flux. The curves are shifted for clarity. (b)
Dependence of the peak-to-peak value $\Delta \alpha$ of the gate
modulation as a function of the external magnetic flux for T1 and
T2. The negative values of $\Delta\alpha$ have the meaning of the
reverse phase-dependencies in plot (a).}
\end{figure}

Gate-modulation curves of sample T1 (low $\lambda$-ratio) for
different values $\varphi_{\rm dc}$  are displayed in
Fig.\,\ref{fig:fig4s}a. We find  a periodic gate-modulation curve
with a modulation depth $|\Delta\alpha|$ of 0.05 rad for
$\varphi_{\rm dc}=0$ and of 0.09 rad for $\varphi_{\rm dc}=\pi$,
whereas for $\varphi_{\rm dc}=\pi/2$, the gate charge sensitivity
disappears almost completely. The gate-dependence period $\Delta
V_{\rm G}$ is roughly 29 mV for both samples and complies with the
value found for the stand-alone test transistor shown in Fig.
\ref{fig:fig2s}b. The gate oscillations are $1e$-periodic, since
the periodicity does not change when applying a magnetic field of
2 T that is sufficient to cause intensive quasiparticle tunneling
\cite{pavolotsky}.

One should note that the gate modulation appears for sample T1
over the whole range of phase $\varphi_{\rm dc}$, hence allowing
efficient qubit control over the whole flux bias range. On the
other hand, for sample T2 ($\lambda=1.9$) the modulation depth is
non-zero only in the vicinity of the point $\varphi_{\rm dc}=\pi$
(cf. Ref.\,\cite{imt2}, where this range was notably smaller) and
almost zero for the rest of the flux-bias range, see Fig.
\ref{fig:fig4s}b.

\begin{figure}
\includegraphics[width=1\linewidth]{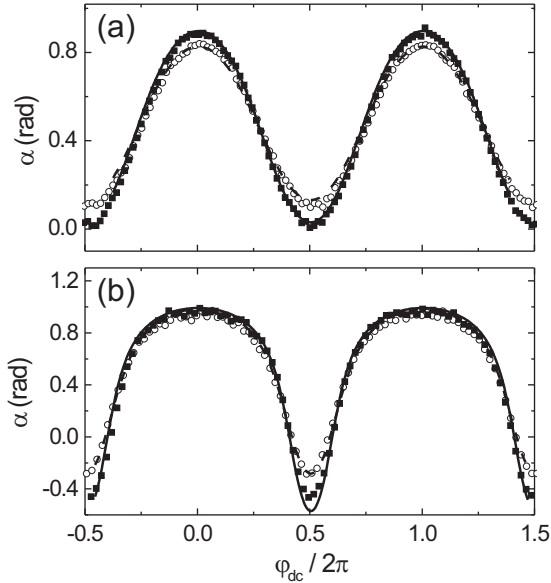}
\caption{\label{fig:fig6} Experimental phase shift for samples (a)
T1 and (b) T2 for two different values of the  gate charge $q=0$
(open circles) and $q=0.5e$ (closed squares). The calculated
dependencies are represented by  solid lines ($q=0.5e$) and dashed
lines ($q=0$), respectively.}
\end{figure}

The  surface plots of the phase shift presented in
Fig.\,\ref{fig:fig5} should reflect the dependence of the local
curvature of the ground-states energy on $q$ and $\varphi$ (see
Eq.\,\ref{curve}). However, because of the finite amplitude of the
oscillations, $a \sim 0.1 a_0^{(1)}$, of phase $\varphi$ these
surfaces yield the values of curvature averaged over the finite
interval. Such averaging of the reverse Josephson inductance is
described by the integral
\begin{equation}\label{average-L}
L_{\rm J}^{-1}(q,\varphi)\rightarrow \frac{1}{\pi}\int_{-1}^1L_{\rm
J}^{-1}(q,\varphi+ax)\frac{dx}{\sqrt{1-x^2}}.
\end{equation}
This expression makes it possible to compare the obtained
experimental data with the corresponding dependencies following
from the theory, taking into account the local curvature of the
ground state energy and the  finite amplitude of the phase
oscillations. By inserting the known parameters $E_{\rm J0}$,
$E_{\rm C}$, $Q$, $\Delta f$, $k$ and $a$ into
Eqs.\,(\ref{detuning}), (\ref{fa}) and (\ref{average-L}), we
calculate the dc bias modulation curves for arbitrary charge on
the transistor gate. The curves for $q=0$ and $q=0.5e$ are shown
in Fig. \ref{fig:fig6}. These curves, which are based on input
data partly deduced from the dc measurements described in Sec. II,
agree well with the primary data from the rf experiments.

\section{\label{sec:level5} Conclusion}

The radio-frequency impedance measurements of the charge-phase
qubit samples with balanced Josephson coupling to Coulomb charging
energy ratio $\lambda$ clearly demonstrated the dependence of the
curvature of the ground state energy on the control charge and
phase in a wide range. The shape of the Josephson inductance
surface of the transistor is well described by the Bloch band
theory.

An advantage of our Nb-technology is, that we were able to
characterize our qubit system at a temperature of $T=4.2$\,K. In
our investigation we found that the Josephson critical current of
the tunnel junctions forming the qubit is scaled with their size,
almost approaching the nominal Ambegaokar-Baratoff value. The
availability of sub-0.1 $\mu$m niobium-based Josephson junctions
was extremely helpful as it offered the valuable new possibility
of studying the influence of thermal fluctuations in an extended
temperature range without increasing the cryogenic efforts. The
experimental results obtained at the large temperature can be
interpreted within a simple quantum-statistical model of the Bloch
transistor.

Although the measured qubit samples T1 and T2 had a charging
energy $E_{\rm C}$ of the island (equal to 80~$\mu$eV and
45~$\mu$eV, respectively) much smaller than the value of the Nb
energy gap, $>$~1~meV, this relation did not ensure the desired
suppression of the quasiparticle tunneling \cite{tuominen}. As
result, instead of the 2\textit{e}-, the 1\textit{e}-periodic
dependence of the qubit Josephson inductance on the gate charge
was observed. Such behaviour of the stand-alone Nb transistor
samples was also observed in earlier measurements performed in dc
configuration, see Ref.\ \cite{dolata05} and references therein.
As was suggested in Ref.\ \cite{dolata05}, this behavior is most
probably due to possible intragap energy states formed in
Nb/AlOx/Nb tunnel barriers or due to non-equilibrium
quasi-particles in the outer electrodes of the transistors
\cite{chargeexp,chargeexp3}. Therefore, further improvement of the
Nb technology of fabrication is required. On the other hand, the
given Nb qubit samples can still operate in the "magic" points
corresponding to the value of the control charge q = 0. The
question as to the rate of the quasiparticle tunneling in the
excited state, presenting the most critical mechanism of the qubit
relaxation \cite{zorin03}, deserves a special study.

\begin{acknowledgments}
We wish to thank H.-P. Duda and R. Harke for valuable technical
assistance and I. Novikov for useful comments on the manuscript.
As well, we would like to acknowledge Th. Weimann and P. Hinze for
their support with the electron-beam writer and B. Egeling and R.
Wendisch for their support in the PECVD and CMP processes. This
work was supported by the European Union (projects SQUBIT-2 and
EuroSQIP).
\end{acknowledgments}


\begin{thebibliography}{00}
\bibitem{comment1} For the parameter range discussed here, the qubit transition
frequency is of the order of 10 GHz in the optimum "magic" points of operation
\cite{cpqb1,zorin03}, which is technically still convenient from the point of
view of driving the qubit by microwave pulses.
\end{thebibliography}

\end{document}